# Ferromagnetism with in-plane magnetization, Dirac spin-gapless semiconducting property, and tunable topological states in two-dimensional rare-earth-metal dinitrides

Yawei Yu,[∥] Xin Chen,[∥] Xiaobiao Liu, Jia Li, Biplab Sanyal, Xiangru Kong,* François M. Peeters, and Linyang Li*

AUTHOR INFORMATION

*Corresponding Authors

   Xiangru Kong − *Center for Nanophase Materials Sciences, Oak Ridge National Laboratory, Oak Ridge, Tennessee 37831, USA;* E-mail: kongx@ornl.gov

   Linyang Li − *School of Science, Hebei University of Technology, Tianjin 300401, China;* E-mail: linyang.li@hebut.edu.cn

Authors

   Yawei Yu − *School of Science, Hebei University of Technology, Tianjin 300401, China*

   Xin Chen − *Department of Physics and Astronomy, Uppsala University, SE-75120 Uppsala, Sweden*

   Xiaobiao Liu − *School of Science, Henan Agricultural University, Zhengzhou 450002, China*

   Jia Li − *School of Science, Hebei University of Technology, Tianjin 300401, China*

   Biplab Sanyal − *Department of Physics and Astronomy, Uppsala University, SE-75120 Uppsala, Sweden*

   François M. Peeters − *Department of Physics and Astronomy, Key Laboratory of Quantum Information of Yunnan Province, Yunnan University, Kunming 650091, China; Department of Physics, University of Antwerp, Groenenborgerlaan 171, B-2020 Antwerp, Belgium*

Author Contributions

[∥]Yawei Yu and Xin Chen contributed equally to this work.

**ABSTRACT**

As the bulk single-crystal $MoN_2/ReN_2$ with a layered structure was successfully synthesized in experiment, transition-metal dinitrides have attracted considerable attention in recent years. Here, we focus on rare-earth-metal (Rem) elements and propose seven stable Rem dinitride monolayers with a 1T structure, namely $1T-RemN_2$. These monolayers have a ferromagnetic ground state with in-plane magnetization. Without spin-orbit coupling (SOC) effect, the band structures are spin-polarized with Dirac points at the Fermi level. Remarkably, the $1T-LuN_2$ monolayer shows an isotropic magnetic anisotropy energy in the *xy*-plane with in-plane magnetization, indicating easy tunability of the magnetization direction. When rotating the magnetization vector in the *xy*-plane, our proposed model can accurately describe the variety of the SOC band gap and two topological states (Weyl-like semimetal and Chern insulator states) appear with tunable properties. The Weyl-like semimetal state is a critical point between the two Chern insulator states with opposite sign of the Chern numbers ($\pm 1$). The large nontrivial band gap (up to 60.3 meV) and the Weyl-like semimetal state are promising for applications in spintronic devices.

**KEYWORDS:** 2D rare-earth-metal dinitrides, ferromagnetic ground state, in-plane magnetization, Weyl-like semimetal, Chern insulator

**Introduction**

Comparing to bulk, the lower dimensional materials have the possibility to generate a multitude of new physical and chemical properties. Currently, we witness a boom in the two-dimensional (2D) material era due to the rapid developments of atomic-thick materials highlighted by graphene[1] and $MoS_2$.[2] Among those 2D materials, transition-metal dinitrides, which exhibit a similar structure as $MoS_2$, have attracted considerable attention in recent years. Bulk single-crystal $MoN_2$/$ReN_2$ with a layered structure can be successfully synthesized experimentally.[3,4] Since the van der Waals interactions between the layers are very weak, $ReN_2$ films have been realized by exfoliating from its bulk material.[4] In theory, $MoN_2$ monolayer was theoretically predicted, where ferromagnetic (FM) and antiferromagnetic (AFM) states correspond to two structural phases.[5,6] In this monolayer, the N-N bond is very important for this magnetic phase transition. By high-throughput calculations, Liu *et al.* proposed a series of monolayers of transition-metal dinitride, such as 1T-$NbN_2$ and 1T-$TaN_2$, which are ferromagnetic.[7] Due to the experimental feasibility and rich magnetic properties of these monolayers, it is thus an interesting question to search for more classes of stable monolayers of transition-metal dinitride.

In 2D FM materials, the appearance of Dirac spin-gapless semiconductors (DSGSs) is very important for spintronics due to the fully spin-polarized band structure and Dirac points at the Fermi level.[8-12] Generally speaking, when considering the spin-orbit coupling (SOC) effect, the Dirac point can be opened to realize a band gap, and then the DSGS can become a Chern insulator, which has attracted extensive study interests due to the realization of the quantum anomalous Hall (QAH) effect.[13,14] Notice that studies of Chern/QAH insulators often assume that the magnetization direction is perpendicular to the plane (out-of-plane) of the 2D system, neglecting the in-plane magnetization.[15] The out-of-plane/in-plane magnetization depends on magnetic anisotropy energy (MAE). Although 2D FM materials with perpendicular magnetic anisotropy have been the subject of numerous studies due to their applications in high-density magnetic recording and spintronic devices,[16,17] the monolayers with in-plane magnetization have the advantage of tunability of the magnetization direction,[18] leading to change in the band structure, which is called as a magneto band-structure effect.[19] This is one of the main motivations to study 2D FM materials with in-plane magnetization.

In this work, we focus on a special class kind of transition-metal elements, rare-earth-metal (Rem) elements (Sc, Y, and Lanthanides), and propose Rem dinitride single layers with a 1T structure, namely 1T-RemN$_2$. Since the 1T-YN$_2$,[20,21] 1T-LaN$_2$,[22] and 1T-GdN$_2$[23] have been investigated in detail, we focus on the other Rem elements. Here, we investigate seven Rem elements (Tb-Lu), and propose the corresponding 1T-RemN$_2$ monolayers, including 1T-TbN$_2$, 1T-DyN$_2$, 1T-HoN$_2$, 1T-ErN$_2$, 1T-TmN$_2$, 1T-YbN$_2$, and 1T-LuN$_2$. By employing first-principles calculations, the structure, dynamical stability, magnetism, and band structure of these monolayers were investigated. Then we used the 1T-LuN$_2$ monolayer as a typical example to investigate the MAE, band structure with SOC, and its tunable topological states.

**Computational method**

The first-principles calculations were performed using the Vienna *ab initio* simulation package (VASP) code,[24-26] implementing density functional theory (DFT). For the electron exchange-correlation functional, we used the generalized gradient approximation (GGA) in the form proposed by Perdew, Burke, and Ernzerhof (PBE).[27] The atomic positions and lattice vectors were fully optimized using the conjugate gradient (CG) scheme until the maximum force on each atom was less than 0.01 eV/Å. The energy cutoff of the plane-wave basis was set to 520 eV with an energy precision of $10^{-6}$ eV in structural optimization, and a higher energy precision of $10^{-8}$ eV was used in other calculations. The Brillouin zone (BZ) was sampled by using a 27×27×1 Γ-centered Monkhorst-Pack grid. The vacuum space was set to at least 20 Å in all the calculations to minimize artificial interactions between neighboring slabs. The phonon spectrum was calculated within the PHONOPY code.[28]

**Structure and stability**

Figure 1(a) shows the optimized crystal structure of 1T-LuN$_2$ monolayer. One Lu atom is bonded with six N atoms to form a stable hexagonal lattice and the space/point group is *P-3m1*/*D$_{3d}$*. As a common structural phase,[29] the 1T structure has been widely studied in theory[7,30] and experiment.[31-33] The optimized lattice constant, Lu-N bond length, and N-N distance along the *z*-axis are 3.64, 2.28, and 1.77 Å, respectively. The dynamical stability of the 1T-LuN$_2$ monolayer is confirmed by its phonon spectrum. As illustrated in Figure 1(b), no imaginary phonon modes are present, confirming the structural stability of 1T-LuN$_2$ monolayer.

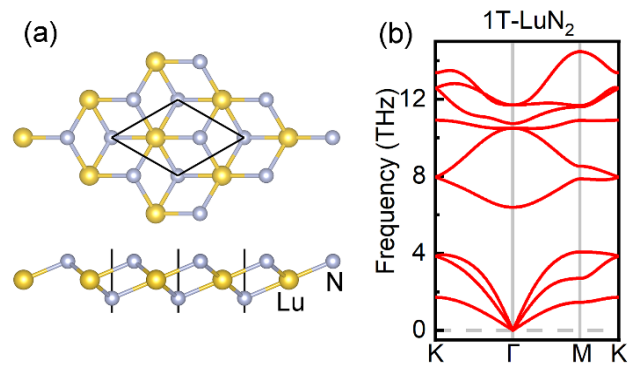

**Figure 1** (a) The optimized crystal structure with top and side views of 1T-LuN$_2$ monolayer. (b) The corresponding phonon spectrum along the high-symmetry path of the first BZ.

The optimized structural parameters for the other six 1T-RemN$_2$ monolayers are summarized in Table 1. For the bond length of Rem-N, these results exhibit the phenomenon of lanthanide shrinkage, where the atomic radius decreases with increasing atomic number. The cohesive energy can be obtained from the equation $E_{coh} = (E_{RemN2} - E_{Rem} - 2E_N)/3$, where $E_{RemN2}$, $E_{Rem}$, and $E_N$ are the total energy of 1T-RemN$_2$ monolayer (per unit cell), of a single Rem atom, and of a single N atom, respectively.[22] The calculated cohesive energy is between −4.33 eV/atom and −4.28 eV/atom, proving the experimental feasibility of 1T-RemN$_2$ monolayers and the chemical similarity of the Rem elements. Additionally, in order to confirm the dynamical stability of the other six 1T-RemN$_2$ monolayers, the phonon spectra are shown in Figure S1 (Part I of Supporting Information). There are no imaginary frequency modes, indicating their dynamical stability. The phonon spectra of the seven 1T-RemN$_2$ monolayers show a similar phonon band structure with low cutoff frequency of acoustic phonons, indicating that these structures are expected to show a low lattice thermal conductivity.[34,35] This property makes 1T-RemN$_2$ monolayer a promising thermoelectric material.

Table 1 The optimized structural parameters of the 1T-RemN$_2$ monolayers. The $a$, $l$, and $h$ are the lattice constant, bond length of Rem-N, and N-N distance along the $z$-axis, respectively. The unit of the cohesive energy $E_{coh}$ is eV/atom.

| 1T-RemN$_2$ | $a$ (Å) | $l$ (Å) | $h$ (Å) | $E_{coh}$ |
|---|---|---|---|---|
| 1T-TbN$_2$ | 3.78 | 2.36 | 1.77 | −4.31 |
| 1T-DyN$_2$ | 3.75 | 2.34 | 1.77 | −4.29 |
| 1T-HoN$_2$ | 3.73 | 2.33 | 1.77 | −4.32 |
| 1T-ErN$_2$ | 3.71 | 2.32 | 1.77 | −4.28 |
| 1T-TmN$_2$ | 3.69 | 2.30 | 1.77 | −4.33 |
| 1T-YbN$_2$ | 3.66 | 2.29 | 1.77 | −4.32 |
| 1T-LuN$_2$ | 3.64 | 2.28 | 1.77 | −4.32 |

**Magnetic property**

To investigate the magnetic ground state, five initial magnetic configurations in 2×2 supercell were considered, including ferromagnetic (FM), Néel antiferromagnetic (NAFM), stripy antiferromagnetic (SAFM), zigzag antiferromagnetic (ZAFM), and nonmagnetic (NM) states.[15] For 1T-LuN$_2$ monolayer, the spin-polarized electron density of the FM/NAFM/SAFM/ZAFM state is shown in Figure S2 (Part II of Supporting Information). The total energy of the NAFM/SAFM/ZAFM/NM state is 295/209/196/390 meV per unit cell relative to that of the FM state, which indicates that the magnetic ground state is FM for 1T-LuN$_2$ monolayer. The 2×2 supercell with the FM state exhibits 12 $\mu_B$ magnetic moment (3 $\mu_B$ per unit cell), and most of the magnetic moment is contributed by the N atoms, as shown in Figure S2(a). By comparing the total energy of the FM and AFM (NAFM/SAFM/ZAFM) states with magnetic moment of $M \approx 1.5\mu_B$ per N atom, the three exchange parameters $J_1 = 8.6$ meV, $J_2 = 1.5$ meV, and $J_3 = 2.4$ meV can be obtained, which correspond to the first, second and third nearest-neighbor magnetic exchange interactions, respectively.[20] By mean-field approximation (MFA),[36] the Curie temperature ($T_c$) can be estimated. Since $M \approx 1.5\mu_B$ is not an integer, two cases of $M = 1\mu_B$ and $M = 2\mu_B$ should be considered respectively. For $M = 2\mu_B$, the partition function can be written as

$$Z = \sum_{m=-2,0,2} \exp[\frac{(\gamma_1 J_1 + \gamma_2 J_2 + \gamma_3 J_3)m\langle M \rangle}{k_B T}],$$

where $\gamma_1 = 3$, $\gamma_2 = 6$, and $\gamma_3 = 3$ are the first, second, and third nearest-neighbor coordination numbers of the N atom, respectively. The average magnetic moment $\langle M \rangle$ is

$$\langle M \rangle = \frac{1}{Z} \sum_{m=-2,0,2} m \exp[\frac{(\gamma_1 J_1 + \gamma_2 J_2 + \gamma_3 J_3)m\langle M \rangle}{k_B T}].$$

It is the critical ratio $(\gamma_1 J_1 + \gamma_2 J_2 + \gamma_3 J_3)/k_B T$ that corresponds to the $T_c$. For $M = 2\mu_B$, this ratio is 3/8, corresponding to $T_c = 1300$ K. For $M = 1\mu_B$, this ratio becomes 1, corresponding to $T_c = 487$ K. Therefore, the estimated Curie temperature by MFA is 487 K < $T_c$ (MFA) < 1300 K. The calculated results of total energy for the FM/NAFM/SAFM/ZAFM/NM states of the other six 1T-RemN$_2$ monolayers are summarized in Table 2, which indicates that the magnetic ground state is FM (magnetic moment of 3 $\mu_B$ per unit cell) for all the 1T-RemN$_2$ monolayers.

**Table 2** Total energy (meV per unit cell) of NAFM/SAFM/ZAFM/NM state with respect to $E$(FM). The unit of MAE determined by the equation $E(001) - E(100)$ and the two fitting coefficients $K_1$ and $K_2$ is μeV per unit cell.

| 1T-RemN$_2$ | $E$(NAFM) | $E$(SAFM) | $E$(ZAFM) | $E$(NM) | MAE | $K_1$ | $K_2$ |
| --- | --- | --- | --- | --- | --- | --- | --- |
| 1T-TbN$_2$ | 409 | 273 | 292 | 513 | 460 | 400.0 | 59.67 |
| 1T-DyN$_2$ | 389 | 262 | 275 | 493 | 536 | 481.2 | 54.34 |
| 1T-HoN$_2$ | 368 | 250 | 259 | 472 | 602 | 551.8 | 50.47 |
| 1T-ErN$_2$ | 350 | 240 | 243 | 453 | 669 | 619.9 | 48.86 |
| 1T-TmN$_2$ | 331 | 229 | 227 | 431 | 707 | 662.0 | 45.16 |
| 1T-YbN$_2$ | 310 | 217 | 209 | 407 | 907 | 842.4 | 64.15 |
| 1T-LuN$_2$ | 295 | 209 | 196 | 390 | 860 | 811.6 | 48.60 |

Based on the FM ground states, we further calculated their MAE. MAE is defined as the required energy to rotate the magnetization direction from the easy axis to the hard axis for a magnetic material,[37] which can be described as MAE = $E$(hard axis) − $E$(easy axis).[16] It is widely accepted that the MAE correlates with the thermal stability of magnetic data storage. Generally, the larger the MAE, the more stable the magnetization and thus the better the performance for data storage.[38] To calculate the MAE of the FM ground state, we included the SOC and calculated total energies of 1T-LuN$_2$ monolayer with magnetization vector $\hat{m}$ constrained in different directions in the *zx*-plane, *zy*-plane, and *xy*-plane. The azimuth angle $\theta$ in the *zx*-plane/*zy*-plane is the angle between the magnetization vector $\hat{m}$ and the positive direction of *z*-axis (001-direction) while the horizontal azimuth angle $\varphi$ in the *xy*-plane is the angle between the magnetization vector $\hat{m}$ and the positive direction of *x*-axis (100-direction). The MAE in the *zx*-plane/*zy*-plane is calculated from the equation $MAE(\theta) = E(001) - E(\theta)$ while the MAE in the *xy*-plane is obtained as $MAE(\varphi) = E(001) - E(\varphi)$. The MAE as a function of $\theta/\varphi$ ranging from 0° to 360° is shown in Figure 2, and $MAE \geq 0$ can be found. In Figure 2(a) and (b), the angle of $\theta = 0°/180°$ corresponds to the smallest MAE while the angle of $\theta = 90°/270°$ corresponds to the largest MAE, proving the *x*-axis/*y*-axis should be the easy axis. For the *xy*-plane (Figure 2(c)), the energy difference between maximum and minimum is less than 0.5 μeV per unit cell, implying essential isotropy and therefore it is an easy plane (*xy*-plane). Hence, 1T-LuN$_2$ monolayer has in-plane magnetization with isotropic MAE in the *xy*-plane, similar as found previously for the VS$_2$ monolayer.[39]

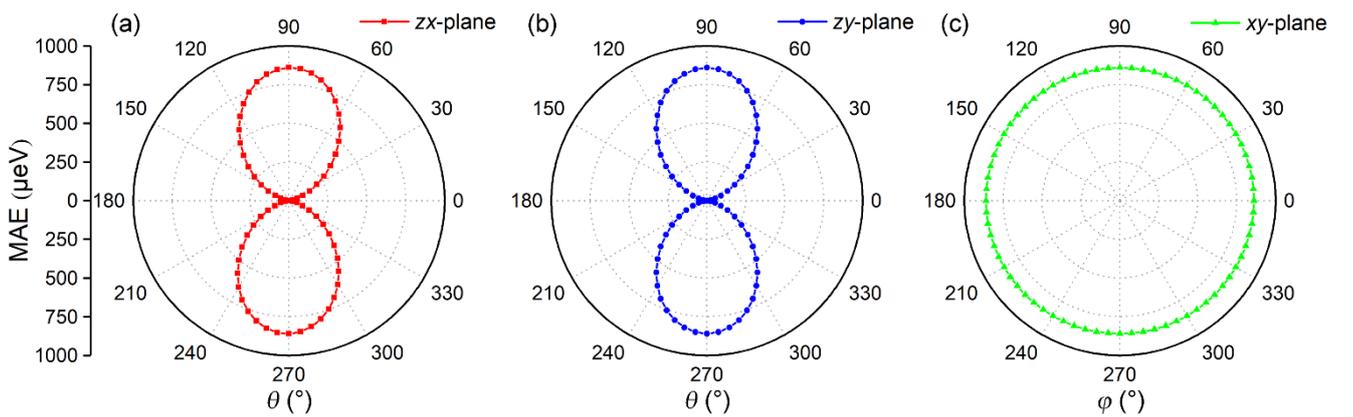

**Figure 2** MAE (1T-LuN$_2$) as a function of the azimuthal angle $\theta$ between magnetization direction $\hat{m}$ and the positive direction of *z*-axis (001-direction) in the *zx*-plane (a) and *zy*-plane (b). MAE (1T-LuN$_2$) as a function of the azimuthal angle $\varphi$ between magnetization direction $\hat{m}$ and the positive direction of

$x$-axis (100-direction) in the $xy$-plane (c).

However, not all FM monolayers with in-plane magnetization exhibit isotropic MAE in the $xy$-plane. For the theoretical OsCl$_3$/PtCl$_3$ monolayer, it was found that the total energy for the different magnetization directions in the $xy$-plane can be anisotropic.[15,18] Since the 1T-LuN$_2$ monolayer exhibit isotropic MAE in the $xy$-plane, the MAE of our hexagonal crystal can be fitted as $\mathrm{MAE}(\theta) = K_1 \sin^2\theta + K_2 \sin^4\theta$.[23,40] Based on our DFT results of $zx$-plane (bule squares in Figure 3(h)), we can obtain the two fitting coefficients, $K_1$ = 811.6 µeV per unit cell and $K_2$ = 48.60 µeV per unit cell. The fitting curve (red line in Figure 3(h)) agrees well with our DFT results and the maximum value of MAE corresponding to $\theta = 90°$ (100-direction) is 860 µeV per unit cell, indicating its considerable stability of the in-plane magnetization. For the 1T-TbN$_2$/1T-DyN$_2$/1T-HoN$_2$/1T-ErN$_2$/1T-TmN$_2$/1T-YbN$_2$ monolayer, the MAE of DFT results (blue squares) and fitting curve (red line) as a function of $\theta$ ($zx$-plane) are shown in Figure 3(b)/(c)/(d)/(e)/(f)/(g). Similar to the results of 1T-LuN$_2$ monolayer in Figure 3(h), all of them exhibit a hard axis along the $z$-axis ($\theta = 0°/180°$) while the easy axis is along 100-direction ($\theta = 90°$). For each 1T-RemN$_2$ monolayer, the maximum value of MAE ($\theta = 90°$, 100-direction) and the fitting coefficients ($K_1$ and $K_2$) are summarized in Table 2. It can be concluded that all the seven 1T-RemN$_2$ monolayers exhibit in-plane magnetization. However, for another Rem element, La,[22] the situation is opposite by using the same equation $\mathrm{MAE}(\theta) = E(001) - E(\theta)$. In Figure 3(a), $\mathrm{MAE} \leq 0$ represents out-of-plane magnetization for 1T-LaN$_2$ monolayer (the easy axis is along $z$-axis), which is similar to the experimental FM monolayers, such as CrI$_3$[41] and Fe$_3$GeTe$_2$.[42]

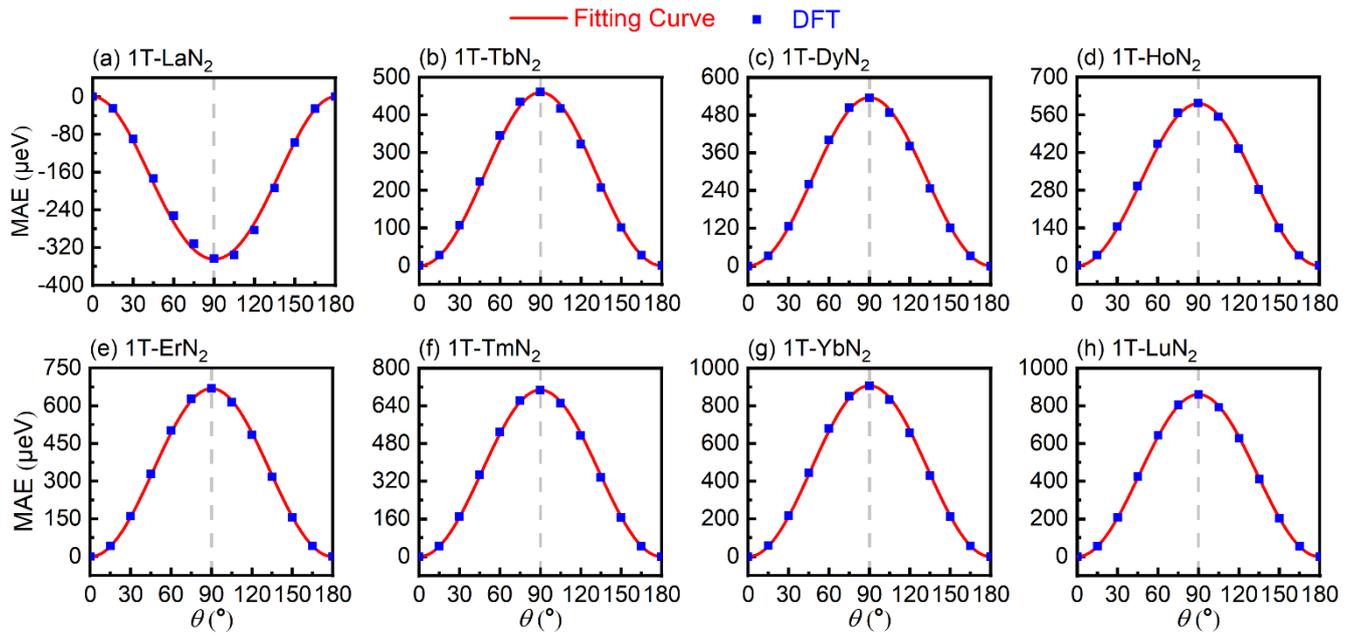

**Figure 3** MAE as a function of azimuth angle $\theta$ in the *zx*-plane for 1T-LaN$_2$ (a)/1T-TbN$_2$ (b)/1T-DyN$_2$ (c)/1T-HoN$_2$ (d)/1T-ErN$_2$ (e)/1T-TmN$_2$ (f)/1T-YbN$_2$ (g)/1T-LuN$_2$ (h) monolayer. The blue squares are from DFT calculations and the red lines are the fitting curves.

**Band structure**

Next, the electronic band structures of the seven 1T-RemN$_2$ monolayers were calculated. In Figure 4, the spin-polarized band structures without SOC are demonstrated. The blue/red bands correspond to the spin-up/spin-down channels, and a giant spin-splitting can be observed. In the spin-down bands, there is a linear Dirac point (insert of Figure 4(g)) at the D point of the first BZ, which is the intersection of the valence and conduction bands at the Fermi level. For the hexagon of the first BZ, D is along the high-symmetry path M-K, and there are twelve Dirac points, as shown in Figure S3(a) (Part III of Supporting Information). To further understand the formation of the Dirac points, the projected band structures of 1T-LuN$_2$ monolayer for the different atoms and atomic orbitals (without SOC) are shown in Figure S4 (Part IV of Supporting Information). For the contributions of the atoms, (a) and (b) of Figure S4 illustrate that the energetic states near the Fermi level are mainly contributed by the nitrogen atoms, while the contributions of the Lutetium atoms are very small. Furthermore, the atomic orbital contributions of the nitrogen atoms are shown in (c), (d), (e) and (f) of Figure S4. The bands near to the Fermi level including the Dirac point are mainly contributed by the $p$ ($p_x$, $p_y$ and $p_z$) atomic orbital, rather than the $s$ atomic orbital. In other words, the 1T-LuN$_2$ monolayer should be a Dirac spin-gapless semiconductor of $p$-state.[9-12]

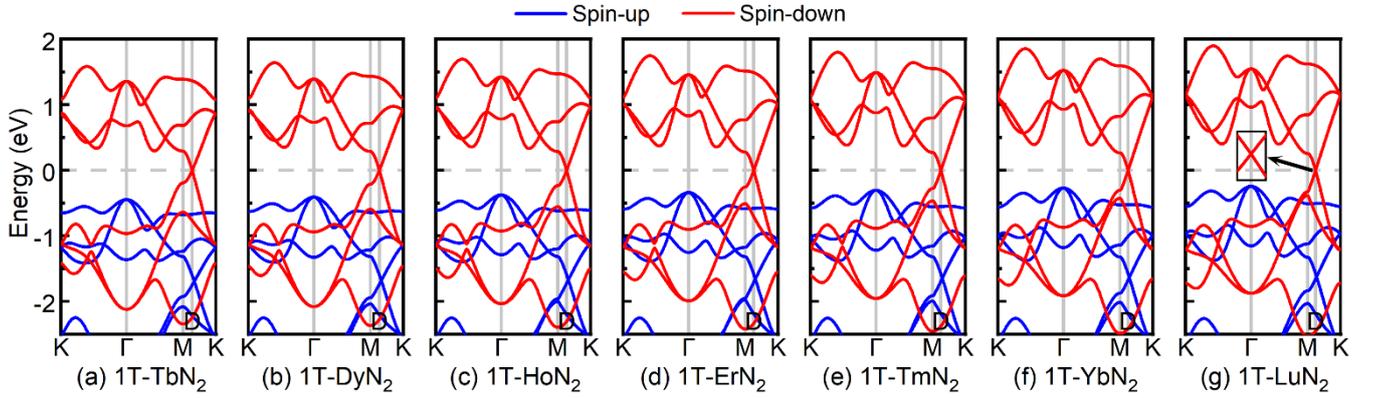

**Figure 4** Band structures without SOC of the seven 1T-RemN$_2$ monolayers.

For the FM monolayers with out-of-plane magnetization, such as CrI$_3$,[41] Fe$_3$GeTe$_2$,[42] and FeB$_3$,[43] the easy axis is along the $z$-axis, leading to the uniqueness of the band structure with SOC. In view of the MAE for 1T-LuN$_2$ monolayer, each direction in the $xy$-plane should be the easy magnetization direction due to the isotropic MAE in the $xy$-plane. Hence, the band structure with SOC along each direction in the $xy$-plane should be studied in theory. When SOC is not included, the Dirac point is located along the high-symmetry path M-K, and there are twelve Dirac points in the hexagon of the first BZ, as shown in Figure S3 (Part III of Supporting Information). We considered the SOC effect for this Dirac band structure of 1T-LuN$_2$ monolayer, and a further study was performed by using WannierTools package.[44,45] Near the Fermi level, the electronic bands are mainly from $p_x$, $p_y$, and $p_z$ atomic orbitals of the N atoms (Figure S4, Part IV of Supporting Information). Thus, we constructed an effective tight-binding Hamiltonian with the three atomic orbitals.[46] By the concept of effective principle layers, an iterative procedure to compute the Green's function for a semi-infinite system was performed,[47] from which the edge states can be obtained.

We first considered the case of $\varphi = 0°$. The blue arrow in Figure 5(a) is the direction of magnetization vector $\hat{m}$, which is $\varphi = 0°$ with respect to the positive direction of *x*-axis. When SOC is included, a band gap is opened at each Dirac point (Figure 5(b)). The global band gaps (between the two red lines) in the path K$_1$-K$_2$/K$_2$-K$_3$/K$_4$-K$_5$/K$_5$-K are 60.3 meV, corresponding to the angle $\alpha = 30°$ between $\hat{m}$ and K$_1$K$_2$/K$_2$K$_3$/K$_4$K$_5$/K$_5$K. The band gaps between the two blue lines in the path K-K$_1$/K$_3$-K$_4$ are 121.7 meV, corresponding to the angle $\alpha = 90°$ between $\hat{m}$ and KK$_1$/K$_3$K$_4$. Here, we should notice $60.3 \approx 121.7 \times \sin 30°$. The calculated chiral edge states for $\varphi = 0°$ are shown in Figure 5(c), and three edge states can be seen. However, only one edge state around the X|X' connects the valence and conduction band areas, confirming that the global band gap of 60.3 meV is nontrivial. This large nontrivial band gap makes it possible to achieve the QAH effect at room temperature. On the other hand, according to the bulk-edge correspondence, one chiral edge state corresponds the Chern number of *C* = +1, which can be further confirmed by Wannier charge centers (WCCs) and anomalous Hall conductivity (AHC) (Figure S5(a) and (b), Part V of Supporting Information). For the case of $\varphi = 30°$ (Figure 5(d) and (e)), $\hat{m}$ is parallel to K$_2$K$_3$/K$_5$K ($\alpha = 0°$), and the global band gaps in the path K$_2$-K$_3$/K$_5$-K vanish ($0 = 121.7 \times \sin 0°$), leading to four gapless points at the Fermi level. Since each gapless point formed by two linear band lines is robust against SOC and twofold degenerate,[48,49] these points can be recognized as Weyl-like points (Part VI of Supporting Information).[50-54] The band gaps between the two blue lines in the path K-K$_1$/K$_1$-K$_2$/K$_3$-K$_4$/K$_4$-K$_5$ are 105.1 meV ($105.1 \approx 121.7 \times \sin 60°$), corresponding to the angle $\alpha = 60°$ between $\hat{m}$ and KK$_1$/K$_1$K$_2$/K$_3$K$_4$/K$_4$K$_5$. Correspondingly, the obtained edge states are shown in Figure 5(f), and we observe a clear "Fermi arc" connecting the pair of gapless points. For the case of $\varphi = 60°$ (Figure 5(g) and (h)), the values of the band gaps are similar to the one for $\varphi = 0°$. The global band gaps (between the two red lines) in the path K-$K_1$/$K_2$-$K_3$/$K_3$-$K_4$/$K_5$-K are 60.3 meV (60.3 ≈ 121.7 × sin30°), corresponding to the angle $\alpha = 30°$ between $\hat{m}$ and K$K_1$/$K_2K_3$/$K_3K_4$/$K_5$K. The band gaps between the two blue lines in the path $K_1$-$K_2$/$K_4$-$K_5$ are 121.7 meV (121.7 = 121.7 × sin90°), corresponding to the angle $\alpha = 90°$ between $\hat{m}$ and $K_1K_2$/$K_4K_5$. Correspondingly, the obtained edge states are shown in Figure 5(i). One edge state connects the valence and conduction band areas, illustrating that the band gap of 60.3 meV is also nontrivial. However, comparing the edge states of $\varphi = 0°$ (Figure 5(c)), it is counter propagating for the case of $\varphi = 60°$, which indicates that the Chern number has opposite sign ($C = -1$), which can be further confirmed by WCCs and AHC (Figure S5(d) and (e), Part V of Supporting Information).

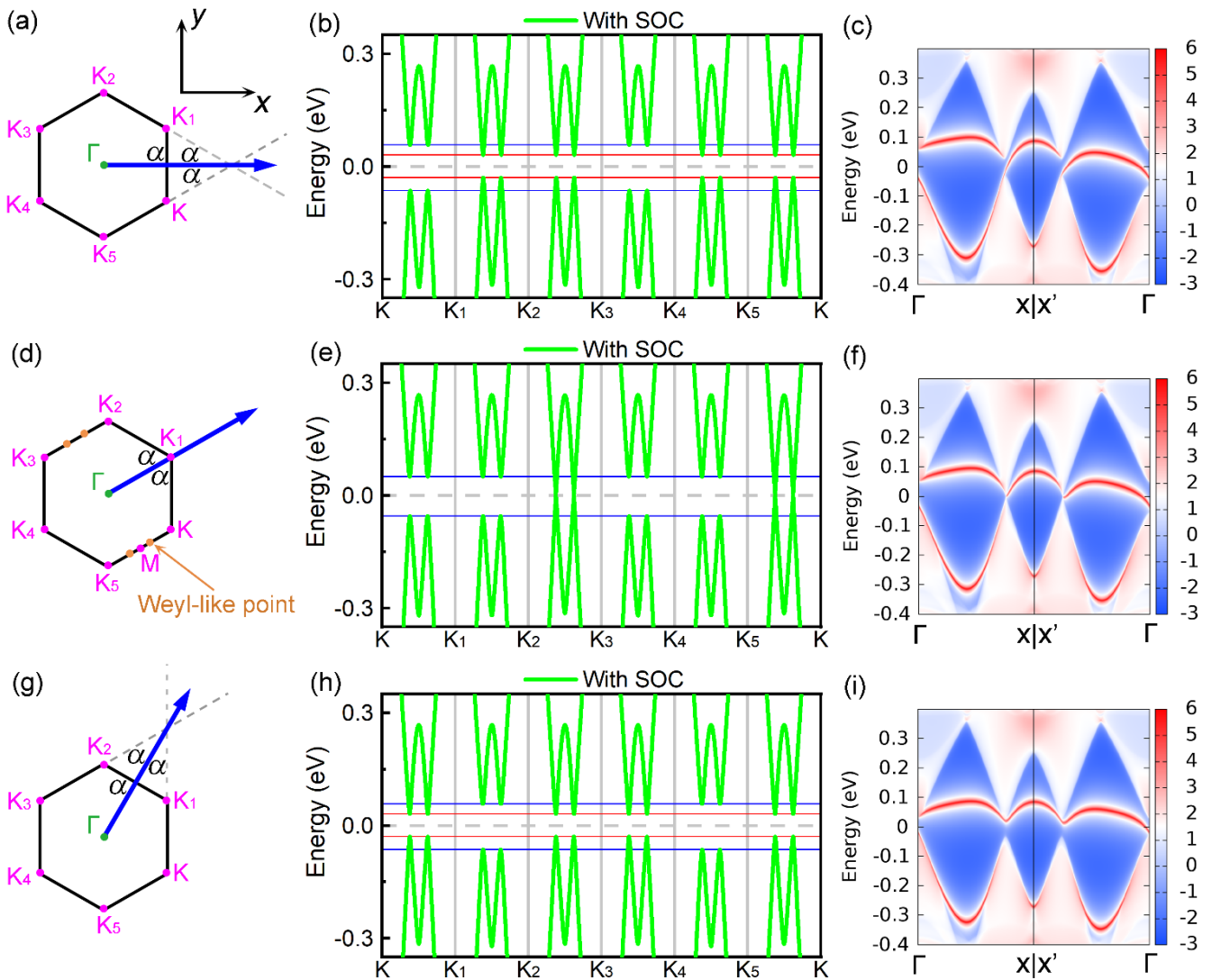

**Figure 5** The direction of magnetization vector $\hat{m}$ (blue arrow) with respect to the positive direction of $x$-axis with angle $\varphi = 0°$ (a), $\varphi = 30°$ (d), and $\varphi = 60°$ (g). The band structure with SOC along the high-symmetry path of K-K$_1$-K$_2$-K$_3$-K$_4$-K$_5$-K (b)/(e)/(h) and the edge states (c)/(f)/(i) for 1T-LuN$_2$ monolayer. (b)(c), (e)(f), and (h)(i) correspond to $\varphi = 0°$ (a), $\varphi = 30°$ (d), and $\varphi = 60°$ (g), respectively.

In the above calculations of the Chern numbers, WCCs are defined as

$$\overline{x}_n(k_y) = \frac{i}{2\pi} \int_{-\pi}^{\pi} dk_x \langle u_n(k_x, k_y) | \partial_{k_x} | u_n(k_x, k_y) \rangle,$$

where $|u_n(k_x, k_y)\rangle$ is the periodic part of Bloch function.[55] By tracking the evolution of the sum of hybrid WCCs, Chern number of $C = +1$ can be observed by the upward shift of WCCs (Figure S5(a), Part V of Supporting Information) in the case of $\varphi = 0°$, while Chern number of $C = -1$ can be observed by the downward shift of WCCs (Figure S5(d)) in the case of $\varphi = 60°$. Moreover, AHC was also calculated according to

$$\sigma_{xy} = \frac{e^2}{(2\pi)^2 h} \int_{BZ} dk_x dk_y f_n(k_x, k_y) \Omega_{n,z}(k_x, k_y),$$

where $f_n(k_x, k_y)$ is Fermi-Dirac distribution function, and $\Omega_{n,z}(k_x, k_y)$ is Berry curvature.[56] As shown in Figure S5(b)/(e), there is a plateau of AHC in the nontrivial SOC band gap of bulk state, which satisfies $\sigma_{xy} = Ce^2/h$, further confirming the Chern number of $C = +1/C = -1$. In order to further understand the topological phase transitions, Berry curvatures are demonstrated in the reciprocal space in Figure S5(c) and (f). As discussed in 1T-YN$_2$ monolayer (out-of-plane magnetization along 001 direction),[21] the total three pairs of Berry curvature peaks with the same positive sign in the whole BZ contribute the Chern number of $C = +3$. In the case of 1T-LuN$_2$ monolayer (in-plane magnetization with $\varphi = 0°$), one pair of Berry curvature peaks flips its sign to be negative as shown in Figure S5(c), leading to $C = +1$. In the case of $\varphi = 60°$, two pairs of Berry curvature peaks flip its sign to be negative as shown in Figure S5(f), leading to $C = -1$.

Essentially, $\alpha\,(0° \leq \alpha \leq 90°)$ is the angle between $\hat{m}$ and the $C_2$ symmetry axis. Due to the $C_{3z}$ symmetry, there are three $C_2$ symmetry axes, $C_{2y}$ (|| $KK_1/K_3K_4$), $C_{2+}$ (|| $K_1K_2/K_4K_5$), and $C_{2-}$ (|| $K_2K_3/K_5K$), as shown in Figure 6(a). When $\hat{m}$ is parallel to the $C_{2-}/C_{2y}/C_{2+}$ axis, the $C_2$ symmetry can be protected, leading to two pairs of Weyl-like points in the corresponding parallel high-symmetry paths.[57-62] When there is an angle $\alpha$ between $\hat{m}$ and the $C_{2-}/C_{2y}/C_{2+}$ symmetry axis, the $C_2$ symmetry can be broken, leading to the absence of Weyl-like points with the opening of a band gap in the corresponding parallel high-symmetry paths. The value of the band gap ($G$) depends on the degree of the broken $C_2$ symmetry, and it can be accurately described by $G = 2g \times \sin\alpha$, where $2g$ is equal to 121.7 meV, corresponding to the value of the maximum SOC band gap in the high-symmetry path K-$K_1$/$K_3$-$K_4$ ($\alpha = 90°$) in the case of $\varphi = 0°$, and g is also equal to the value of the global SOC band gap in the case of $\varphi = 0°$. Using this model, the value of the SOC band gap on each high-symmetry path can be calculated, and then the global SOC band gap can be obtained for each angle of $\varphi$ in the $xy$-plane, as shown in Figure 6(b) (red lines). To confirm the rationality of this model, we calculated the global SOC band gaps by DFT, as shown in Figure 6(b) (blue dots). Notice that the two results agree very well, indicating the accuracy of our proposed model.

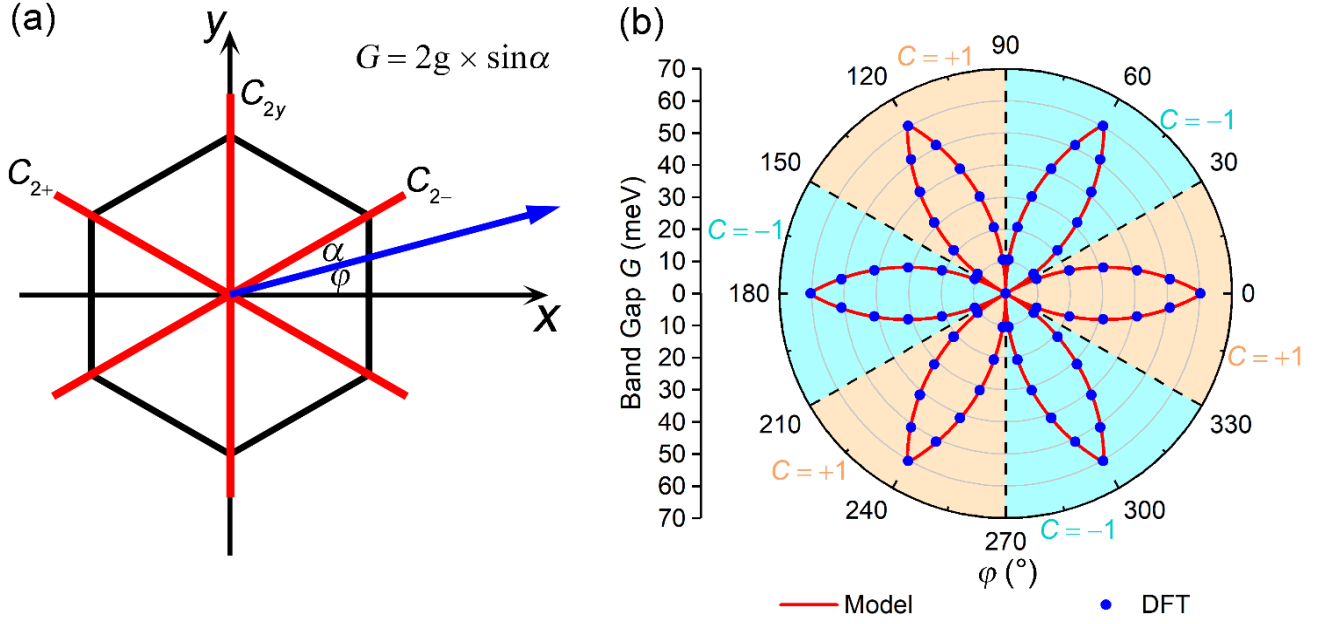

**Figure 6** (a) $\alpha$ is the angle between magnetization vector $\hat{m}$ (blue arrow) and the $C_2$ symmetry axis. $\varphi$ is the angle between magnetization vector $\hat{m}$ and the positive direction of x-axis. (b) Global SOC band gap $G$ of the 1T-LuN$_2$ monolayer as a function of the azimuthal angle $\varphi$ (xy-plane) based on the proposed model (red lines) and DFT calculations (blue dots). The different color areas correspond to different Chern numbers ($C = \pm 1$), and the black dashed lines between the different color areas represent the Weyl-like semimetal states ($G = 0$).

Thus, we have proven that the Weyl-like semimetal state is a critical point between two Chern insulator states with opposite sign of the Chern numbers ($C = \pm 1$), leading to tunable topological states when rotating the magnetization vector in the xy-plane. By breaking time reversal symmetry and protecting mirror (glide mirror) symmetry, the gapless points under SOC can be realized in other monolayers.[18,53,63-69] In our 1T-LuN$_2$ monolayer, where time reversal symmetry is broken and $C_2$ symmetry is protected, the 2D Weyl-like points can be realized in special magnetization directions of xy-plane ($\varphi = 30°/90°/150°/210°/270°/330°$). Moreover, a tunable Chern number is also important. Experimentally, a well quantized QAH effect with tunable Chern number (up to $C = 5$) in multilayer structures has been realized.[70] Theoretically, it has been proven that the irradiation of left/right circularly polarized light can tune the Chern number.[71] In the present work, when the magnetization vector is rotated in the $xy$-plane, it can realize a change of sign of the Chern number $C$, corresponding to a change of propagating direction of the edge channel, making 1T-LuN$_2$ monolayer promising for applications in spintronics.

**Conclusion**

In summary, by DFT calculations, we predict a series of stable 2D rare-earth-metal dinitrides, 1T-RemN$_2$ monolayers, which exhibit FM ground state (magnetic moment of 3 $\mu_B$ per unit cell). Without SOC, all the considered systems show a spin-polarized electronic band structure, and are Dirac spin-gapless semiconductor. Taking SOC into account, an in-plane magnetization can be found in all the 1T-RemN$_2$ monolayers with a maximum MAE in the range 460~907 μeV per unit cell, indicating the considerable stability of the in-plane magnetization. For the 1T-LuN$_2$ monolayer, the in-plane magnetization exhibits isotropic MAE in the $xy$-plane, indicating easy tunability of the magnetization direction. By rotating the magnetization direction in the $xy$-plane, the value of the nontrivial SOC band gap can be accurately described by our proposed model. The gapped band structures (up to 60.3 meV) correspond to the Chern insulator state while the gapless band structures correspond to the Weyl-like semimetal state, which is a critical state between two Chern insulator states with opposite sign of the Chern numbers ($C = \pm 1$). The realization of Chern insulator and Weyl-like semimetal states in the 1T-LuN$_2$ monolayer has therefore enriched the family of topological materials.

**Supporting Information**

Phonon spectra (Part I), Spin-polarized electron density for 1T-LuN$_2$ (Part II), Positions of the Dirac

point and band structure without SOC of 1T-LuN$_2$ (Part III), Projected band structures without SOC of 1T-LuN$_2$ (Part IV), Chern numbers and Berry curvatures (Part V), and Weyl-like points (Part VI).

**Acknowledgements**

This work is supported by the National Natural Science Foundation of China (Grant No. 12004097), the Natural Science Foundation of Hebei Province (Grant No. A2020202031), and the Foundation for the Introduced Overseas Scholars of Hebei Province (Grant No. C20200313). X.L. acknowledges financial support from the National Natural Science Foundation of China (Grant No. 22005087). The computational resources utilized in this research were provided by Shanghai Supercomputer Center. A portion of this work (X.K.) was conducted at the Center for Nanophase Materials Sciences which is a US Department of Energy Office of Science User Facility. This research used resources of the Compute and Data Environment for Science (CADES) at the Oak Ridge National Laboratory, which is supported by the Office of Science of the U.S. Department of Energy under Contract No. DE-AC05-00OR22725.

**Notes**